\documentclass[aps,twocolumn,showpacs,superscriptaddress,pra,10pt]{revtex4-2}

\usepackage{graphicx}
\usepackage{dcolumn}
\usepackage{bm}
\usepackage{hyperref}

\IfFileExists{newtxtext.sty}
    {\usepackage{newtxtext,newtxmath}}
    {\IfFileExists{stix.sty}
       {\usepackage{stix}}
       {\IfFileExists{mathptmx.sty}
       {\usepackage{mathptmx}}{} } }

\usepackage[dvipsnames]{xcolor}

\begin{document}

\preprint{APS/123-QED}

\title{
Spectrum of density, spin and pairing fluctuations of an attractive two-dimensional Fermi gas
}

\author{Christian Apostoli}
\email{christian.apostoli@unimi.it}
\affiliation{Dipartimento di Fisica ``Aldo Pontremoli'', Universit\`a degli Studi di Milano, via Celoria 16, 20133 Milano, Italy}

\author{Patrick Kelly}
\affiliation{Institute for Physical Science and Technology, University of Maryland, College Park, Maryland 20742, USA}

\author{Annette Lopez}
\affiliation{Department of Physics, Brown University, 182 Hope Street, Providence, RI 02912, USA}

\author{Kaelyn Dauer}
\affiliation{Department of Physics, California State University Fresno, Fresno, California 93720, USA}

\author{Gianluca Bertaina}
\affiliation{Istituto Nazionale di Ricerca Metrologica, Strada delle Cacce 91, I-10135 Torino, Italy}

\author{Davide Emilio Galli}
\email{davide.galli@unimi.it}
\affiliation{Dipartimento di Fisica ``Aldo Pontremoli'', Universit\`a degli Studi di Milano, via Celoria 16, 20133 Milano, Italy}

\author{Ettore Vitali}
\email{evitali@mail.fresnostate.edu}
\affiliation{Department of Physics, California State University Fresno, Fresno, California 93720, USA}

\date{\today}

\begin{abstract}
We leverage random phase approximation and unbiased auxiliary-field quantum Monte Carlo methods
to compute
dynamical correlations for a dilute homogeneous two-dimensional attractive Fermi gas. Our main purpose is to quantitatively study the collective excitations of the system to generate robust benchmark results and to shed light into fermionic superfluidity in the strongly correlated regime. In particular we are motivated by a recent paper suggesting that 
the Higgs mode can be detected in the spectrum of spin fluctuations.
Despite the fact that we are somewhat limited by finite-size effects, our study  pinpoints indeed a clear peak in the spin channel at low momentum, but a detailed analysis suggests that such a peak, though certainly interesting, does not correspond to the Higgs mode. We propose a different explanation for the shape of the spin structure factor.
On the other hand, our results clearly show that the Higgs mode can be detected in the density channel at very small wave vectors, although very good resolution is necessary. 
\end{abstract}

\maketitle

\section{\label{sec:intro}Introduction}

The {\it ab-initio} calculation of the dynamical correlation functions of a strongly-correlated quantum system is a fundamental and challenging task which gives access to critical information about the system's excitation spectrum. In particular, these correlation functions can be used to probe collective excitations, such as the Nambu-Goldstone mode and the more elusive celebrated Higgs or amplitude mode in superfluids.
The Higgs mode is particularly hard to observe and it has been the center of a long-standing experimental effort, both in condensed matter physics~\cite{Shimano_HiggsModeSuperconductors_2020,Pekker_AmplitudeHiggsModes_2015} and in atomic physics.
In the realm of ultracold bosons, the Higgs mode has been investigated in a series of experiments employing lattice shaking or cavity-enhanced Bragg spectroscopy~\cite{Bissbort_DetectingAmplitudeMode_2011,Endres_Higgsamplitudemode_2012,Leonard_MonitoringmanipulatingHiggs_2017}.
In ultracold Fermi gases, the actual detection of the amplitude mode has proved particularly challenging, since the Higgs mode in these systems has a very slow-decaying spectral tail~\cite{Cea_NonrelativisticDynamicsAmplitude_2015,Kurkjian_Linearresponsesuperfluid_2020}. This corresponds to real-time damping due to the proliferation of quasiparticle excitations. Only very recently, experimental observations were reported, which used direct coherent excitation of the mode with radiofrequency pulses, Bragg spectroscopy following an interaction quench, or interaction modulation via periodic tuning of the magnetic field~\cite{Behrle_Higgsmodestrongly_2018,Bayha_Observingemergencequantum_2020,Dyke_HiggsOscillationsUnitary_2024,kell2024exciting}.

On the theoretical and computational side, the Higgs mode in neutral Fermi systems has been studied with various methodologies, including e.g. the time-dependent Bogoliubov-de Gennes equations~\cite{ScottRamps2012,HannibalQuench2015,Tokimoto_ExcitationHiggsMode_2019,Lyu_ExcitinglonglivedHiggs_2023}, time-dependent density functional theory~\cite{Barresi_GenerationdecayHiggs_2023}, the functional-integral method~\cite{Phan_FollowingHiggsmode_2023,Han_ObservabilityHiggsmode_2016,Cea_NonrelativisticDynamicsAmplitude_2015}, and exact diagonalization (for few-body systems)~\cite{Bjerlin_FewBodyPrecursorHiggs_2016}. More recently, relying on the generalized random-phase approximation, it has been suggested that the Higgs mode 
of an attractive fermionic system could be detected in the dynamical correlations of the spin degrees of freedom~\cite{Zhao_2020}, 
which can be measured experimentally
through spin-sensitive Bragg spectroscopy~\cite{HoinkaDynamicSpinResponse2012,Vale_Spectroscopicprobesquantum_2021,Senaratne_Spinchargeseparationonedimensional_2022}. Such measurements would extend the insightful investigations that can be performed in the density fluctuations sector~\cite{Hoinka_Goldstonemodepairbreaking_2017,Biss_ExcitationSpectrumSuperfluid_2022,Sobirey_ObservingInfluenceReduced_2022,Dyke_HiggsOscillationsUnitary_2024}. Ref.~\cite{Zhao_2020} also argued that the two-dimensional (2D) configuration, as contrasted to the three-dimensional one mostly considered so far, renders the Higgs signal in the spin sector particularly prominent.

The investigation of spin correlation functions to address collective modes, and in particular the Higgs mode, in Fermi superfluids is fascinating, but correctly interpreting the results may be subtle. The key advantage of a spin probe is that the correlation functions at low momentum are expected to be significantly different from zero only close to $2\Delta$ ($\Delta$ being the superfluid gap), which is exactly where the Higgs mode is expected, while the lower energy Goldstone mode is, at least partially, filtered out. 
The difficulty, on the other hand, is to 
discriminate whether and to what extent the Higgs mode can indeed be excited with a spin probe. 
The analysis in \cite{HE2016470} in fact suggests that beyond mean-field studies are necessary to detect the coupling between the Higgs mode and the spin fluctuations. 

Our plan for this work is to gain further insight by systematically running generalized random phase approximation calculations (GRPA)~\cite{Ganesh_Collectivemodes_2009,Kelly_OntheaccuracyofRPA_2021,Zhao_2020,PhysRevA.80.063627,PhysRevA.74.042717} for the  the dynamical structure factor $S({\bf{q}},\omega)$, its spin homologous $S^s({\bf{q}},\omega)$, and the pairing amplitude dynamical structure factor $S^a({\bf{q}},\omega)$ (defined below). In addition, we leverage unbiased quantum Monte Carlo (QMC) results to explore the effect of many-body correlations in the density and the spin channel.

In the density channel, we find qualitative agreement between the QMC results and the GRPA spectrum, with a renormalization of the Goldstone mode.
Our GRPA calculations, for very large lattices and at very small momentum (not accessible with QMC), also suggest that the Higgs mode can be detected in the density dynamical structure factor.
In the spin channel, 
the GRPA results show a very interesting peak at low momentum, consistent with the results in \cite{Zhao_2020}, and the correlated QMC results appear to confirm the existence of such a peak, with a renormalized energy. However, our GRPA analysis, relying on $S^a({\bf q},\omega)$, indicates that 
this spin peak cannot be interpreted as the Higgs mode and we propose a different kinematic explanation, rooted in the density of states of pairs of quasiparticles. 

The paper is organized as follows: Section~\ref{sec:methods} explains the Hamiltonian and the methodologies that we used: GRPA, QMC and the analytic continuation procedure. Then in Section~\ref{sec:results} we describe our results and present a systematic comparison between GRPA and QMC. Finally, we draw our conclusions and present some perspectives in Section~\ref{sec:conclusions}.

\section{\label{sec:methods} Model and Methods}
Our starting point is the Hamiltonian for a collection of attractive unpolarized spin-$1/2$ fermions of equal mass $m$ moving in 2D. In this context, the two spin species, denoted by $\uparrow$ and $\downarrow$, respectively, represent two hyperfine states of $^6\mathrm{Li}$ or $^{40}\mathrm{K}$ atoms. We focus on the dilute regime, where the fine details of the inter-particle forces can be neglected, and consider an attractive zero-range interaction $v_{\uparrow\downarrow}({\bf{r}}_1,{\bf{r}}_2) = -g \delta({\bf{r}}_1-{\bf{r}}_2)$, $g>0$.
The Hamiltonian can be written, in field-theoretical notation, as:
\begin{equation}
\label{methods:H}
    \hat{H} = \hat{K} + \hat{V} 
\end{equation}
with:
\begin{equation}
    \hat{K} = \int d{\bf{r}} \, \sum_{\sigma} \hat{\psi}_{\sigma}^{\dagger}({\bf{r}})
    \, \left( - \frac{\hbar^2 \nabla^2}{2m} \right)  \hat{\psi}^{}_{\sigma}({\bf{r}})
\end{equation}
and:
\begin{equation}
   \hat{V} = -g \int d{\bf{r}} \, \hat{\psi}_{\uparrow}^{\dagger}({\bf{r}}) \, \hat{\psi}_{\downarrow}^{\dagger}({\bf{r}}) \, \hat{\psi}_{\downarrow}^{}({\bf{r}})
   \, \hat{\psi}_{\uparrow}^{}({\bf{r}}) 
\end{equation}
Our main objective is the calculation of the zero-temperature spin and density dynamical structure factors of the system in the dilute regime, in order to extract information about the low-energy collective excitations.
We explore the dependence on the interaction strength $\log(k_F a)$, where $k_F$ is the Fermi momentum (which will be defined later) and $a$ the $s$-wave scattering length (in this work, we adopt the convention that the dimer binding energy $\epsilon_b$ of the contact model is related to $a$ by $|\epsilon_b|=\hbar^2/ma^2$~\cite{Bertaina_Twodimensionalshortrangeinteracting_2013}). 

The zero-temperature density dynamical structure factor of $N$ fermions is defined as the following two-body dynamical correlation function:
\begin{equation}
\label{methods:sqo}
    S({\bf{q}},\omega) = \int_{0}^{+\infty} \frac{dt}{2\pi N} \, e^{i \omega t} \, \left\langle e^{i \hat{H} \frac{t}{\hbar}} \, \hat{n}_{\bf{q}} \, e^{- i \hat{H} \frac{t}{\hbar}}
    \, \hat{n}_{-{\bf{q}}}
    \right\rangle
\end{equation}
where the angular brackets denote a ground state expectation value and $\hat{n}_{\bf{q}}$ is the Fourier component of the particle density operator $\hat{n}({\bf{r}}) = \hat{n}_{\uparrow}({\bf{r}}) + \hat{n}_{\downarrow}({\bf{r}}) = \hat{\psi}_{\uparrow}^{\dagger}({\bf{r}})\hat{\psi}_{\uparrow}^{}({\bf{r}}) 
+ \hat{\psi}_{\downarrow}^{\dagger}({\bf{r}})\hat{\psi}_{\downarrow}^{}({\bf{r}})$.
Similarly, we define the spin dynamical structure factor $S^s({\bf{q}},\omega)$ 
by replacing, in \eqref{methods:sqo}, the total particle density with the spin density $\hat{n}^s({\bf{r}}) = \frac{1}{2} \left( \hat{n}_{\uparrow}({\bf{r}})
- \hat{n}_{\downarrow}({\bf{r}}) \right)$.

We address the singularities due to the contact potential in \eqref{methods:H} through lattice regularization, i.e., by introducing a high-momentum cutoff in the kinetic energy term:
\begin{equation}
\label{methods:KL}
   \hat{K} \simeq  \sum_{\sigma}\int_{[-\pi/b,\pi/b)^2} d{\bf{k}} \,  \varepsilon(k) \,
     \hat{\psi}_{\sigma}^{\dagger}({\bf{k}})  \hat{\psi}_{\sigma}^{}({\bf{k}})
\end{equation}
where the dispersion relation is, as usual, $\varepsilon(k)= \frac{\hbar^2 k^2}{2m}$. In Eq.~\eqref{methods:KL}, we integrate over the first Brillouin zone of a square lattice with parameter $b$, $(b\mathbb{Z})^2$. In addition, we further regularize the problem by introducing a supercell $\mathcal{L} = [-L/2,L/2]^2 \cap (b\mathbb{Z})^2$, i.e., a finite square lattice with $M= L/b \times L/b$ sites, and choosing periodic boundary conditions (PBC). The choice of PBC restricts the integration in \eqref{methods:KL} to a finite summation over the set of allowed momenta ${\bf{k}} = \frac{2\pi}{L} {\bf{n}}$, with ${\bf{n}} \in \mathbb{Z}^2 $ such that ${\bf{k}} \in [-\pi/b,\pi/b)^2$.
We observe that, within this regularization technique, the continuum limit can be recovered by letting $b \to 0$ and the infinite system limit can be recovered by letting $L \to +\infty$.
The interaction term in the Hamiltonian is regularized by introducing a contact interaction of the form $v_{\uparrow\downarrow}({\bf{r}}_1,{\bf{r}}_2) = -U b^{-2} \delta_{{\bf{r}}_1,{\bf{r}}_2}$, where $\delta$ is now the discrete Kronecker delta, while the interaction strength of the discrete model is tuned in such a way that the lattice two-body problem has the same zero-energy $s$-wave scattering length $a$ as the problem in the continuum. 
This regularization procedure maps the original Hamiltonian into a lattice model which we can write as:
\begin{equation}
\label{methods:HL}
    \hat{H}_{\mathcal{L}} = \sum_{{\bf{k}},\sigma} \varepsilon(k) \, \hat{\psi}_{\sigma}^{\dagger}({\bf{k}})  \hat{\psi}_{\sigma}^{}({\bf{k}})
    - U \sum_{{\bf{r}}} \hat{n}_{\uparrow}({\bf{r}})
    \hat{n}_{\downarrow}({\bf{r}})
\end{equation}
Using the typical notation for ``Hubbard-like'' Hamiltonians, we can write $\varepsilon(k) = t |b {\bf{k}}|^2$, with hopping amplitude given by $t=\frac{\hbar^2}{2mb^2}$, and, as shown in \cite{PhysRevA.92.033603}:
\begin{equation}
    \frac{U}{t} = \frac{4\pi}{\log(k_F a) + \log(\mathcal{C} \sqrt{n})}
\end{equation}
where $n=N/M$ is the particle density on the lattice, $k_F = \frac{\sqrt{2\pi n}}{b}$ is Fermi momentum, $a$ is the $s$-wave scattering length of the original problem in the continuum, while $\mathcal{C}= 0.80261$ is a constant. In the following we will denote the Fermi energy of the lattice model by $\varepsilon_F = \varepsilon(k_F)$.

The crucial advantage of stepping from \eqref{methods:H} to \eqref{methods:HL} is the possibility, for attractive unpolarized fermions, irrespective of the total density (filling), to calculate exactly the intermediate scattering function in imaginary time:
\begin{equation}
    \label{Methods:Ftau}
    F({\bf{q}},\tau) = \frac{1}{N} \left\langle e^{ \tau \hat{H}_{\mathcal{L}} } \, \hat{n}_{\bf{q}} \, e^{- \tau \hat{H}_{\mathcal{L}}
}    \, \hat{n}_{-{\bf{q}}}
    \right\rangle, \quad \tau \geq 0
\end{equation}
both in the density and the spin channels (substituting $\hat{n}$ with $\hat{n}^s$), by leveraging unbiased auxiliary-field quantum Monte Carlo methodologies (AFQMC), which are extensively explained in \cite{Zhang_AFQMC_2013,PhysRevB.94.085140,JCP10.1063/1.4861227}. In fact, it is possible to map the imaginary time evolution operator $\exp(- \tau \hat{H}_{\mathcal{L}})$ into a random walk in the manifold of $N$ particles' Slater determinants, modeling independent fermions moving in a stochastic external field (the ``auxiliary field''). It has been shown that, whenever the system is spin-balanced and the interaction is attractive, the infamous sign problem does not affect the calculations and we can find exact results with a computational time which is polynomial in the size of the system \cite{PhysRevA.92.033603,PhysRevB.94.085140}. The dynamical structure factors can then be obtained by performing analytic continuation of the imaginary time data~\cite{vitali_initio_2010,Bertaina_Statisticalcomputationalintelligence_2017,PhysRevA.102.053324}.

We employ the differential evolution for analytic continuation (DEAC) algorithm~\cite{PhysRevE.106.025312}, which evolves a population of candidate solutions for the dynamical structure factor over several generations, each one improving their average fitness and adaptively adjusting the control parameters. We use the DEAC code by A. Del Maestro's group~\cite{DEAC_code}. To assess the robustness of our results, we also performed some cross checks (not shown) employing the genetic inversion by falsification of theories (GIFT) algorithm~\cite{vitali_initio_2010,Bertaina_Statisticalcomputationalintelligence_2017}.

\begin{figure*}
\includegraphics{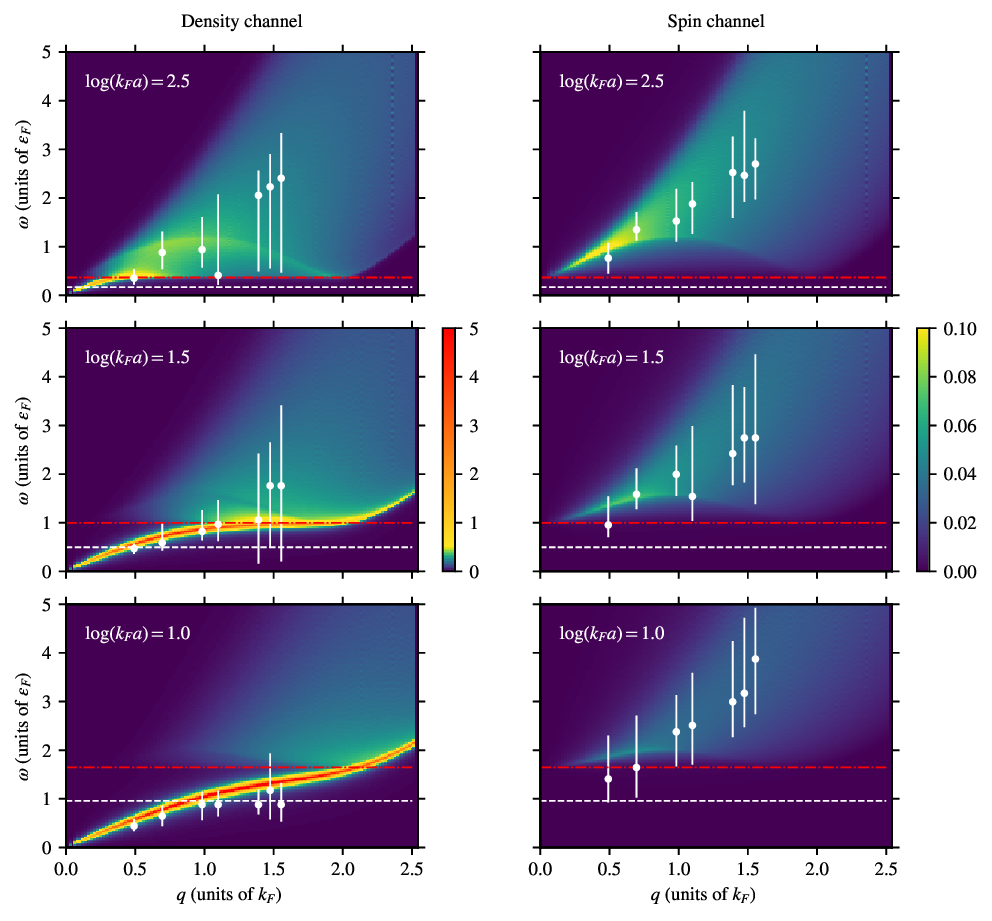}
\caption{\label{fig:colormaps} Dynamical structure factors (arbitrary units) in the density (left) and spin (right) channels for interaction strengths $\log (k_F a) = 2.5$ (top), $1.5$ (middle), and $1.0$ (bottom). Dots: maxima of the structure factors obtained by QMC for a system of $26$ fermions, with vertical bars representing their full width at half maximum. Color plots: dynamical structure factors (arbitrary units) obtained by GRPA for a larger system of $N=5850$ fermions at the same density. Dashed horizontal lines: twice the quasiparticle gap for the system of $N=26$ fermions; for interaction strength $\log (k_F a) = 2.5$, it is calculated from the theory by Gor'kov and Melik-Barkhudarov~\cite{PhysRevA.67.031601,SovPhysJETP.Gorkov}, which was shown to be accurate in this weak-interaction condition~\cite{PhysRevA.96.061601}; for interaction strengths $\log (k_F a) = 1.5$ and $1.0$, it is calculated with QMC from Ref.~\cite{PhysRevA.96.061601}. Dash-dotted horizontal lines: twice the quasiparticle gap predicted by BCS theory for the system of $N=5850$ fermions.}
\end{figure*}
\begin{figure*}
\includegraphics{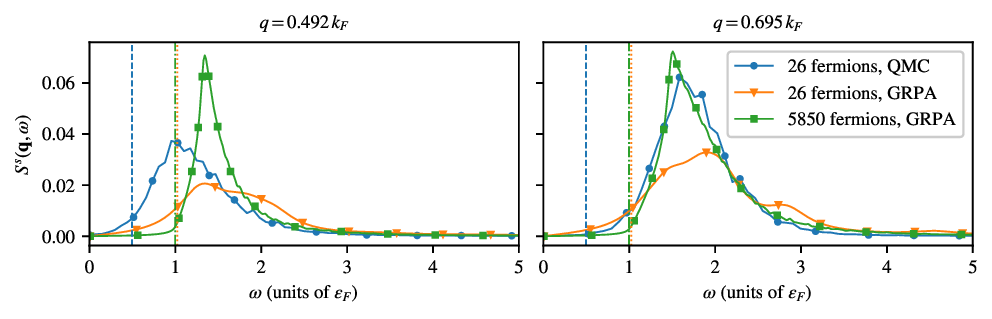}
\caption{\label{fig:spin_logkFa1.5} Spin dynamical structure factor $S^s({\bf{q}},\omega)$ (arbitrary units) for interaction strength $\log (k_F a) = 1.5$ and wave numbers $q = 0.492\,k_F$ (left panel) and $0.695\,k_F$ (right panel). The solid lines represent our results, with markers indicating selected data points to improve readability. Blue dots: QMC results for 26 fermions. Orange triangles: GRPA results for 26 fermions. Green squares: GRPA results for 5850 fermions. Dashed vertical line: twice the quasiparticle gap for 26 fermions, calculated with QMC from Ref.~\cite{PhysRevA.96.061601}. Dotted vertical line: twice the quasiparticle gap predicted by BCS theory for 26 fermions. Dash-dotted vertical line: twice the quasiparticle gap predicted by BCS theory for 5850 fermions.}
\end{figure*}

In addition to unbiased QMC calculations, we compute the dynamical structure factors within the generalized random phase approximation (GRPA)~\cite{Ganesh_Collectivemodes_2009,Kelly_OntheaccuracyofRPA_2021,Zhao_2020,PhysRevA.80.063627,PhysRevA.74.042717}, still relying on the above lattice regularization. GRPA implements a linear response theory approach, studying the response of the local density to a time-dependent perturbation coupled to the density itself, resulting in a Hamiltonian of the form:
\begin{equation}
\hat{H}(t) = \hat{H}_{\mathcal{L}} - \mu \hat{N} + \sum_{r} u({\bf r},t) \hat{n}({\bf r}) 
\end{equation}
where $\hat{H}_{\mathcal{L}}$ is the lattice Hamiltonian \eqref{methods:HL}, $\mu$ is a chemical potential, $\hat{N}$ the particle number operator, while $u({\bf r},t)$ is an external potential. An analogous definition holds for the spin perturbations.
GRPA approximately computes the response function $\chi({\bf{q}},\omega)$, yielding the change in the local density as the system evolves unitarily under the influence of the perturbation:
\begin{equation}
\label{methods:lin_res_theory}
\delta \langle \hat{n} \rangle ({\bf{q}},\omega) =\chi({\bf{q}},\omega) u({\bf{q}},\omega)
\end{equation}
where the angular brackets denote an expectation value with respect to the time dependent quantum state, evolving from the unperturbed ground state. Within GRPA, the Hamiltonian $\hat{H}_{\mathcal{L}} - \mu \hat{N}$ is replaced by the most general mean-field breakup \cite{Kelly_OntheaccuracyofRPA_2021}, and the order parameters are, at each time instant, self-consistently adjusted to follow the time-dependent perturbation. Finally, the dynamical structure factors can be obtained through the fluctuation-dissipation theorem:
\begin{equation}
\label{methods:sqo_chi_relation}
S({\bf{q}},\omega) = - \frac{\hbar}{\pi n} \lim_{\eta \to 0^{+}} 
 \chi''({\bf{q}},\omega + i \eta) 
\end{equation}
where $n$ is the average density, $\chi''=\Im \chi$, and $\eta$ is a convergence parameter. In actual GRPA numerical implementations, the parameter $\eta$ has to be set to some small positive number, as we will discuss in Section~\ref{sec:results}.
Our GRPA computations allow us to make direct comparison with QMC and with previous results in \cite{Zhao_2020}, where a different regularization is used. In addition, GRPA helps us perform a mode-coupling analysis to explore the suggestion in \cite{Zhao_2020} to use the spin dynamical structure factor to detect the elusive Higgs mode. 
At the same time, since GRPA calculations only scale linearly with the lattice size $M$ (they just require numerical summations over the Brillouin zone), their results allow us to readily estimate size effects in the calculations.

\section{\label{sec:results}Results}
\begin{figure*}
\includegraphics{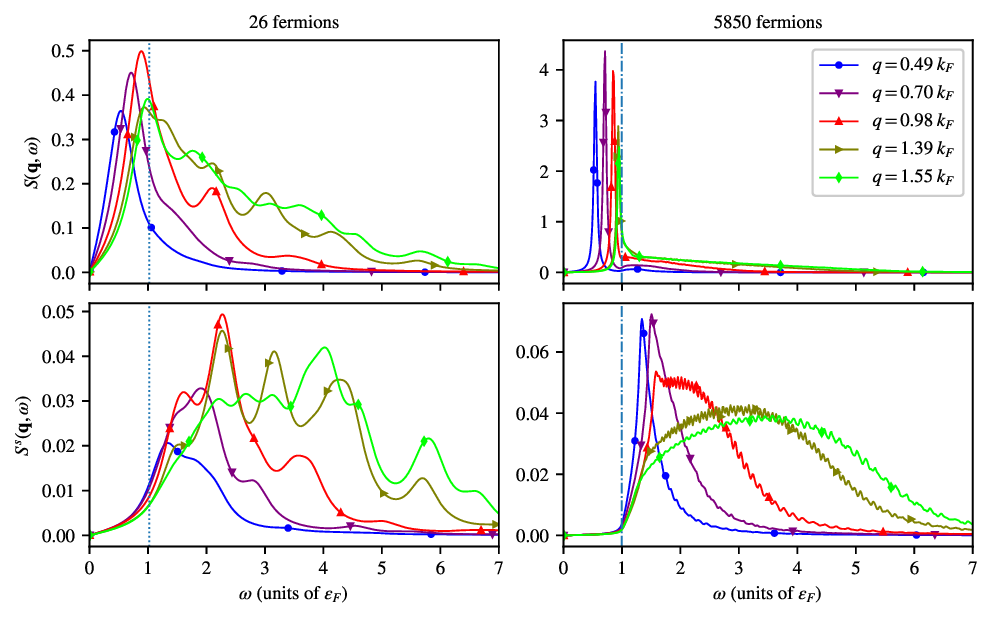}
 \caption{\label{fig:GRPA_GRPA_comparison} GRPA results for the dynamical structure factors (arbitrary units) in the density (top panels) and spin (bottom panels) channels of a system of 26 (left panels) and 5850 (right panels) fermions, for interaction strength $\log (k_F a) = 1.5$ and various values of the wave number $q$. The solid lines represent our results, with markers indicating selected data points to improve readability. Dotted vertical lines: twice the quasiparticle gap for 26 fermions, as predicted by BCS theory. Dash-dotted vertical lines: twice the quasiparticle gap for 5850 fermions, as predicted by BCS theory.
 }
\end{figure*}
\begin{figure*}
\includegraphics{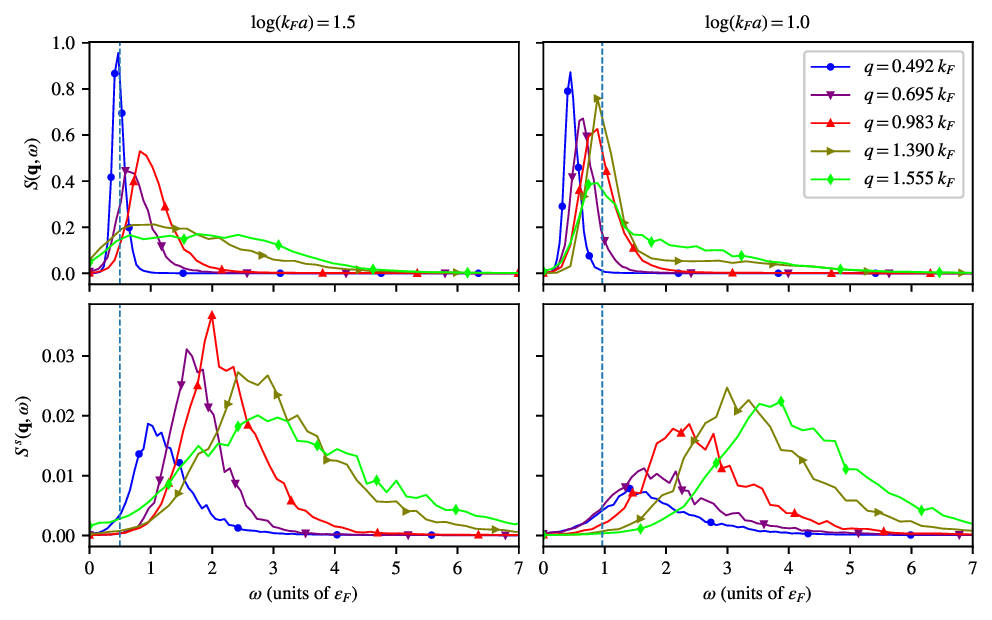}
\caption{\label{fig:QMC_all_k} QMC results for the dynamical structure factors (arbitrary units) in the density (top panels) and spin (bottom panels) channels, for interaction strengths $\log (k_F a) = 1.5$ (left panels) and $1.0$ (right panels), and various values of the wave number $q$. The solid lines represent our results, with markers indicating selected data points to improve readability. Dashed vertical lines: twice the quasiparticle gap, calculated with QMC from Ref.~\cite{PhysRevA.96.061601}.}
\end{figure*}
\begin{figure*}
\includegraphics{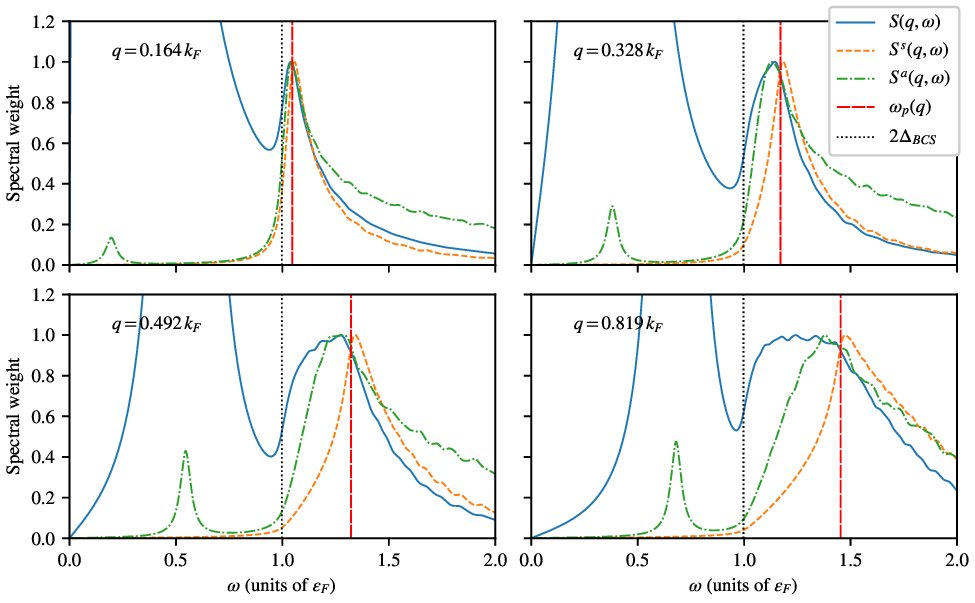}
\caption{\label{fig:Sqw_maxima_comparison_logkFa_1_5} 
GRPA dynamical structure factors (arbitrary units) in the density, spin and amplitude channels, for 5850 fermions at interaction strength $\log (k_F a) = 1.5$ and wave numbers $q=0.164 \, k_F$ (upper left), $q = 0.328 \, k_F$ (upper right), $q = 0.492 \, k_F$ (lower left), and $q = 0.819 \, k_F$ (lower right). For ease of comparison, the dynamical structure factors were scaled so that their peaks in the range $\omega > 2\Delta_{BCS}$ have the same height, conventionally set to $1$. The vertical dotted line marks twice the BCS quasiparticle gap. The vertical dashed line marks $\omega_{p}(q)$, the maximum of the density of pair quasiparticle states of total momentum $q$.}
\end{figure*}
Motivated by the paper \cite{Zhao_2020}, we address two key questions: are there significant discrepancies between GRPA and correlated AFQMC calculations in the description of the density and spin dynamical correlations?
If AFQMC confirms, at least qualitatively, the GRPA picture displaying a peak in the spin channel, were the authors of \cite{Zhao_2020} justified in claiming that such a peak can be identified with the celebrated Higgs mode?

\subsection{Comparison between GRPA and AFQMC}
We use AFQMC to compute exactly the dynamical correlations in imaginary-time \eqref{Methods:Ftau} for a system of $N=26$ atoms using a regularization square lattice with $M=1225$ sites.
In this context, ``exactly'' means that, for the given system size, the systematic error in the properties of the lattice model is below the level of the statistical uncertainty. Thanks to the spin balance and the absence of the infamous sign problem, this can be achieved with polynomial scaling.
Previous AFQMC studies by some of us \cite{PhysRevA.96.061601} on the spectral function of the 2D Fermi gas indicate that the lattice size and number of particles used in this paper are large enough to achieve accurate estimations in the bulk and in the continuum limit. This is also the largest closed-shell size we are able to study with the algorithm for two-body correlations described in \cite{PhysRevB.94.085140}. 
As discussed below, we use GRPA as a tool to further assess finite-size effects.
Our calculations yield the imaginary-time intermediate scattering functions $F({\bf{q}},\tau)$ in both the density and spin channels, for momenta as small as $q \simeq 0.5 k_F$, $k_F$ being the Fermi momentum. 
We perform DEAC analytic continuation of $F({\bf{q}},\tau)$ to extract the density dynamical structure factor $S({\bf{q}},\omega)$ and the spin dynamical structure factor $S^s({\bf{q}},\omega)$. We study the
behavior of the structure factors as we increase the interaction strength from the weakly correlated BCS regime with $\log(k_Fa) = 2.5$ to the more strongly correlated, though still not molecule-dominated, regime with $\log(k_Fa) = 1.0$. The choice of the values of interaction was dictated by the intent to compare with the results in Ref.~\cite{Zhao_2020}.

In Fig.~\ref{fig:colormaps}, our AFQMC results in the density (left column) and spin channels (right column) are shown as circles, representing the maxima of the structure factors, together with vertical bars representing the widths at half heights (the detailed shapes of the AFQMC structure factors are shown in Figs.~\ref{fig:spin_logkFa1.5} and ~\ref{fig:QMC_all_k}). The results are superimposed to background color plots, obtained with GRPA, for $S({\bf{q}},\omega)$ (left column) and for $S^s({\bf{q}},\omega)$ (right column) for a much larger system at the same particle density, precisely $N=5850$ atoms moving on a regularization lattice with ${M=525\times525}$ sites. 
The larger system allows for the investigation of a denser set of momentum values, providing a clearer picture, and gives important information about finite-size effects, as we will discuss below. The horizontal lines represent $2\Delta$, $\Delta$ being the superfluid gap (i.e., the energy needed to break a Cooper pair) as computed within a BCS mean-field approach (say $\Delta_\mathrm{BCS}$), informing the GRPA calculations (red dashed-dotted lines), or within AFQMC (white dashed lines) \cite{PhysRevA.96.061601}.

Our GRPA results, despite the different choice of regularization, appear to be consistent with the results in \cite{Zhao_2020}.
In the density channel, a sharp Nambu-Goldstone mode is visible at low energy $\omega<2\Delta_\mathrm{BCS}$; for higher energy $\omega>2\Delta_\mathrm{BCS}$ the quasiparticle pair continuum emerges, and the Nambu-Goldstone mode is strongly damped. At the AFQMC level, the lattice supercell is large enough to explore the behavior below the superfluid gap, and the results confirm the existence of a sharp Nambu-Goldstone mode, whose dispersion is renormalized by the correlations beyond mean-field: the peak at the lowest values of the momentum is at lower energy with respect to the GRPA prediction (this is not a finite-size effect, as we will show below), consistently with a significantly smaller superfluid gap, as was found in previous calculations in Ref.~\cite{PhysRevA.96.061601}. The AFQMC results at higher momentum are also consistent with the emergence of a quasiparticle pair continuum, as shown by the increasing size of the bars. Incidentally, we observe that the Higgs mode is expected to appear in the density structure factor, although the Goldstone mode has a much higher spectral weight. Indeed, while such an effect is beyond the resolution of the color plot, in Fig.~\ref{fig:Sqw_maxima_comparison_logkFa_1_5} (blue solid line) we show that, at the GRPA level, we see a second peak which, as we will discuss below, can be interpreted as the Higgs mode. On the other hand, at the AFQMC level, the dominant peak of the Goldstone mode and the broadening due to the quasiparticle pair continuum make the requirements for the detection of the amplitude mode beyond the resolution we can achieve with analytic continuation.

In the spin channel, as expected, within GRPA no excitation is present for $\omega<2\Delta_\mathrm{BCS}$. At higher energies, the GRPA spin dynamical structure factor appears to develop a sharp peak at low momentum which broadens at higher momenta. By comparing different interaction strengths, we observe that the spin structure factor gets flatter and fainter as the interaction strength increases (and $\log (k_F a)$ decreases). This is consistent with the system being more and more similar to a Bose-Einstein condensate of tightly bound molecules, where the spin degrees of freedom are suppressed. We notice that the widths of the QMC results increase with the coupling more than within GRPA.

The key result in \cite{Zhao_2020} was the observation 
of the sharp spin mode, say $\omega_s(q)$, which the authors interpreted as the celebrated Higgs mode, with a dispersion of the form $\omega_s(q)=2\Delta_\mathrm{BCS} + \alpha q^2$ as $q/k_F \to 0$.
Our GRPA calculations are also consistent with a sharpening of the dynamical structure factor at low momentum, although we feel the need to be very cautious about the interpretation of this spin ``mode'' as the evidence of a detection of the Higgs mode, as we will discuss in details below. Before diving into this discussion, we comment about the comparison with AFQMC.
At the unbiased AFQMC level, we are somewhat limited by the system size, which does not allow us to compute the structure factors below $|{\bf q}|\simeq 0.5 k_F$. Our AFQMC results for the spin dynamical structure factor are shown in the right column of Fig.~\ref{fig:colormaps} and, in more detail, in the subsequent Figs.~\ref{fig:spin_logkFa1.5} and \ref{fig:QMC_all_k}.
At the lowest considered momenta, the shape of the spin dynamical structure factor displays a clear peak, whose intensity decreases with the interaction strength and whose energy may be compatible with a mode that would converge to $2\Delta$ (with the AFQMC pairing gap) in the limit $q \to 0$.
At the same time, the peak does not appear very sharp, in particular if we compare it with the peaks in the density channel. Nevertheless, we suggest that the broadening of the peak may be due to the finite size. In order to corroborate such a statement, 
in Fig.~\ref{fig:spin_logkFa1.5} we show the spin dynamical structure factor $S^s({\bf{q}},\omega)$ for interaction strength $\log (k_F a) = 1.5$, focusing on the two smallest wave numbers $q = 0.492\,k_F$ and $0.695\,k_F$.
We compare the AFQMC spectra for $26$ atoms with the GRPA ones for $26$ and $5850$ atoms at the same density. All of them show a peak that moves to higher frequency as $q$ increases from $0.492\,k_F$ to $0.695\,k_F$. The GRPA spectra exhibit maxima at approximately the same positions for both system sizes, with a sharper peak for the larger system. 
The AFQMC results have a width which is compatible with the corresponding GRPA result for $26$ atoms, while the peaks are shifted to lower frequency, as expected from the renormalization of the gap. 
If we assume that the significant sharpening observed at the GRPA level for increasing size also happens at the correlated level, then we can suggest that a sharp mode does indeed exist also at the AFQMC level, and it is not an artifact of GRPA.

We take a closer look at the size dependence within GRPA in Fig.~\ref{fig:GRPA_GRPA_comparison},
but, before delving into it, a discussion of our choices for the convergence parameter $\eta$ is in order. The dynamical structure factor resulting from the GRPA theory is achieved in the limit $\eta \to 0^{+}$, see Eq.~\eqref{methods:sqo_chi_relation}. In our implementation, we choose a finite positive value to broaden Dirac deltas in the dynamical structure factor. We tune the values empirically, and in particular we choose $\eta \simeq 0.06 \varepsilon_F$ for $26$ atoms and $\eta \simeq 0.03 \varepsilon_F$ for $5850$ atoms.

In Fig.~\ref{fig:GRPA_GRPA_comparison}, we compare the GRPA spectra for interaction $\log (k_F a) = 1.5$ and for two system sizes: $N=26$ and $N=5850$ atoms. The larger system shows the same qualitative behavior, and the positions of the peaks do not show significant change.
On the other hand, the larger size displays significantly sharper peaks in both channels, allowing for a clearer distinction between the collective modes and the quasiparticle pair continuum. On one hand, this highlights that in the $26$-atoms system there are still relevant finite-size effects, thus encouraging future efforts to allow AFQMC to access larger sizes; at the same time, if we assume that GRPA provides an accurate recipe for size corrections, this allows us to strengthen our claim that a sharp mode does indeed exist at the QMC level in the spin channel.

Fig.~\ref{fig:QMC_all_k} shows the density (top panels) and spin (bottom panels) dynamical structure factors yielded by our QMC calculations for interaction strengths $\log (k_F a) = 1.5$ and $1.0$ and selected momenta. The qualitative picture presented by the two interaction strengths is similar, the main quantitative difference being due to the different value of the superfluid gap.
The density structure factors for small $q$ are dominated by the Nambu-Goldstone peak. As $q$ increases, this peak moves to higher frequency and becomes less sharp, eventually merging with the quasiparticle pair continuum when it reaches the energy of twice the superfluid gap. The spin dynamical structure factors for small $q$ exhibit a peak above twice the quasiparticle gap.
This peak moves to higher frequency as $q$ increases and eventually significantly broadens and becomes compatible with a quasiparticle pair continuum.

\subsection{ Higgs mode or spin mode?}
We have thus observed that, within the resolution of our calculations, AFQMC qualitatively confirms the GRPA description of the dynamical structure factors. The main quantitative difference appears to be rooted in the unbiased superfluid gap being significantly smaller than the mean-field value, which informs GRPA.
On this ground, we were able to reproduce the interesting peak in the spin channel originally highlighted by \cite{Zhao_2020}. In this section, we wish to leverage our GRPA study to address the claim that such a peak is the Higgs mode. 
The first observation is that, within GRPA, there is no direct coupling between the dynamics of the order parameter and the spin excitations, as it is evident from Eq.~14 in Ref.~\cite{Zhao_2020} and the analysis in Ref.~\cite{HE2016470} (Eq.~120 and following).
More explicitly, let us denote $\chi_{\alpha,\beta}({\bf{q}},\omega)$ the matrix of linear response functions (like the one in \eqref{methods:lin_res_theory}), with labels $\alpha, \beta=n,s,\Delta,\Delta^{\dagger}$, $n$ denoting particle density, $s$ denoting spin density and $\Delta$ being the on-site pairing $\hat{\Delta}({\bf{r}})=\hat{\psi}^{}_{\downarrow}({\bf{r}})\hat{\psi}^{}_{\uparrow}({\bf{r}})$. Within GRPA, we rigorously find that $\chi_{n,s}({\bf{q}},\omega) = \chi_{\Delta,s}({\bf{q}},\omega) = \chi_{\Delta^{\dagger},s}({\bf{q}},\omega) = 0$; in other words, a perturbation coupled to the spin density (a magnetic field) is not able to excite modulations in the particle density or in the superfluid order parameter. 
As a consequence, the relation:
\begin{equation}
    \int_{0}^{+\infty}d\omega \, \chi^{\prime\prime}_{\Delta,s}({\bf{q}},\omega) \propto 
    \langle \Psi_{BCS} | \hat{\Delta}({\bf{q}}) \, \hat{n}^s(-{\bf{q}}) | \Psi_{BCS} \rangle
\end{equation}
(where, again, $\chi^{\prime\prime}$ denotes the imaginary part of $\chi$) implies that
the quantum state $\hat{n}^s(-{\bf{q}}) | \Psi_{BCS} \rangle$ ($| \Psi_{BCS} \rangle$ being the BCS ground state) must be orthogonal to all quantum states of the form $\hat{\Delta}({\bf{q}}) | \Psi_{BCS} \rangle$; in simple words, if we induce a spin modulation on top of the BCS ground state we obtain a quantum state that has no overlap with states which are obtained by modulating the order parameters (Nambu-Goldstone and Higgs).

In Fig.~\ref{fig:Sqw_maxima_comparison_logkFa_1_5} (left panel) we show an example of a GRPA calculation of density dynamical structure factor $S({\bf{q}},\omega)$, spin dynamical structure factor  $S^{s}({\bf{q}},\omega)$, as well as amplitude dynamical structure factor
constructed from the response function of the operator $\hat{\Delta} + \hat{\Delta}^{\dagger}$ (Hermitian part of the pairing operator) as follows:
\begin{equation}
    S^a({\bf{q}},\omega) \propto \chi^{\prime\prime}_{\Delta,\Delta}
+ \chi^{\prime\prime}_{\Delta,\Delta^{\dagger}} +  \chi^{\prime\prime}_{\Delta^{\dagger},\Delta}
+ \chi^{\prime\prime}_{\Delta^{\dagger},\Delta^{\dagger}}
\end{equation}
We focus on wave vectors within the Fermi surface.
At the two smallest wave vectors shown in the figure, we notice that both $S({\bf{q}},\omega)$ (blue solid line) and $S^a({\bf{q}},\omega)$ (green dashed-dotted line) display the Goldstone mode below $2\Delta_{BCS}$ (black dotted vertical line) and higher energy peaks at exactly the same energy above $2\Delta_{BCS}$. The actual definition of $S^a({\bf{q}},\omega)$ as the amplitude response makes it very natural to interpret such peaks as the Higgs mode. As the magnitude of the wave vector increases, the secondary peak in the density channel broadens due to the quasiparticle pair continuum.
On the other hand, the spin dynamical structure factor (orange dashed line) shows a peak but at a slightly higher energy, with a discrepancy increasing with the magnitude of the wave vector. Our conclusion is thus that, while a peak exists in the spin channel, it cannot be interpreted as a manifestation of the amplitude (Higgs) mode of the order parameter.

We propose, on the other hand, that such a spin ``mode'' has a kinematic origin.
This is related to the BCS dispersion relation $E({\bf{k}}) = \sqrt{(\varepsilon(k)-\mu)^2 + \Delta_{BCS}^2}$, which governs the possible quasiparticle excitations. When a spin probe with momentum ${\bf{q}}$ acts on the system, we expect the formation of pairs of quasiparticles with momenta ${-\bf{k}}$ and ${\bf{k}}+{\bf{q}}$ for all possible values of ${\bf{k}}$. Indeed, one can easily check that the peaks in the spin dynamical structure factor follow very closely the maxima $\omega_{p}({\bf{q}})$ of the (suitably $\eta$-regularized) function $\sum_{{\bf{k}}} \delta( \omega - (E({\bf{k}}+{\bf{q}}) + E({-\bf{k}}))$, reported as red dashed vertical lines in Fig.~\ref{fig:Sqw_maxima_comparison_logkFa_1_5}. This observation suggests that the spin structure factor is entirely governed by the quasiparticle pair continuum, which is pretty ``narrow'' at small momentum, giving rise to a peak in the response function.

We comment that the actual possibility to resolve the Higgs mode from the peak in the spin channel required the study of a very large system ($\simeq 6000$ fermions). For smaller systems, finite-size effects broaden the structures (see Fig.~\ref{fig:spin_logkFa1.5}), thus making it impossible to have enough resolution. This also implies that, within our current capabilities, it would not be realistic to investigate this discrepancy using AFQMC. We also notice that, at the AFQMC level we cannot exclude that a coupling between the pairing and spin channel might exist due to correlations beyond mean-field, but any such coupling, if it even exists, would be well below our resolution.

\section{\label{sec:conclusions}Discussion and Conclusions}

In summary, our study leveraged generalized random phase approximation and the unbiased auxiliary-field quantum Monte Carlo method to compute density and spin density dynamical correlation functions for a two-dimensional homogeneous system of attractive fermions. 
Our current AFQMC algorithm allowed us to compute the structure factors, in the $(q,\omega)$ plane, for momenta as low as $0.5 \, k_F$, which turned out to be small enough to resolve the collective Nambu Goldstone mode below the superfluid gap. On the other hand, with GRPA we were able to study much larger systems, which allowed us to assess the severity of finite-size effects, as well as to gain further insight into the physical interpretations of the results.

Incidentally, we observe that the possibility to obtain unbiased two-body dynamical correlations for Fermi superfluids for small momenta, well inside the Fermi sea, is very important beyond the scope of atomic physics, with potential applications in nuclear physics and nuclear astrophysics.

Comparison between AFQMC and GRPA results, together with the analysis of size effects obtained via GRPA, has revealed overall qualitative agreement.
In the density channel a Nambu-Goldstone mode is clearly evident, very sharp at low energy and momentum and merging with a quasiparticle pair continuum at higher energies. The many-body correlations renormalize the dispersion relation of the mode and the superfluid gap, controlling the threshold for the emergence of the quasiparticle pair continuum. In addition, our GRPA results for $S({\bf q},\omega)$ and $S^a({\bf q},\omega)$ in very large systems clearly show that the Higgs mode appears as a secondary peak in the density dynamical structure factor at very small wave vectors. The resolution needed to detect this secondary peak is beyond our capability with AFQMC.

We also observe a distinct peak in the spin channel, which had been previously highlighted in \cite{Zhao_2020} and considered as evidence of the detection of the celebrated Higgs mode. The observation in \cite{Zhao_2020} was an important motivation for our study. 
This peak in the spin channel indeed 
displays a dispersion which is compatible with a quadratic behavior reminiscent of the celebrated Higgs mode. Our analysis, though, implies that, at least within GRPA, such a peak must have an origin which is different from the dynamics of the amplitude of the order parameter, as we extensively discuss. Our suggestion is that the peak is entirely due to the density of states of pairs of quasiparticles, which can be excited when a spin probe acts on the system.
Our QMC results indicate a significant renormalization of the dispersion of such a spin ``mode'' due to the many-body correlations, consistent with a smaller value of the pairing gap in the unbiased calculations with respect to the mean-field result. On the other hand, it appears natural to us to assume that the origin of the spin peak is still rooted in the quasiparticle kinematics. 

Despite the challenges posed by finite-size effects, our study thus contributes to the fundamental understanding of two-dimensional Fermi superfluids, providing unbiased results which can serve as crucial benchmarks for many-body theories and can provide important insight to experimental researchers in the important challenge of measuring collective modes in superfluids.

A particularly exciting direction in the study of collective modes involves the investigation of the role of spin polarization in attractive Fermi gases~\cite{Vitali_ExoticSuperfluidPhases_2022,Radzihovsky_ImbalancedFeshbachresonantFermi_2010,bertaina_density_2009}; in fact, recent numerical findings confirmed the stability of an elusive Fulde-Ferrell-Larkin-Ovchinnikov phase at zero temperature \cite{Vitali_ExoticSuperfluidPhases_2022}, whose experimental detection is very complicated: one of the possibilities is to measure the anisotropy of the speed of sound, which can be extracted from dynamical structure factors. In this context, the availability of numerical results, like the ones we presented in this paper, can be a very important asset to inform and guide the experimental search. For spin imbalanced gases the inclusion of $p$-wave contributions might also be relevant~\cite{Bertaina_QuantumMonteCarlo_2023}.

As we navigate these perspectives, our findings open the doors for theoretical and experimental advancements, pushing the boundaries of our comprehension of dynamical properties within strongly correlated quantum systems.

\section{Acknowledgements}

E.V. acknowledges support from the National Science Foundation, grant number PHY-2207048. 
The AFQMC calculations were performed on the ACCESS (previously XSEDE) resources.
Analytic continuation and GRPA were run on computational resources provided by INDACO Platform, which is a project of High Performance Computing at Universit\`a degli Studi di Milano.

\bibliography{apssamp}

\end{document}